\documentstyle[12pt]{article}

\begin{document}

\begin{titlepage}
\begin{flushright}
IFUP--TH--48/95
\end{flushright}
\vskip 1truecm
\begin{center}
\Large\bf
2--dimensional Regge gravity \\
in the conformal gauge\footnote{This work is  supported in part
  by M.U.R.S.T.}
\end{center}
\vskip 1truecm
\begin{center}
{Pietro Menotti and Pier Paolo Peirano} \\
{\small\it Dipartimento di Fisica dell'Universit\`a, Pisa 56100,
Italy and}\\
{\small\it INFN, Sezione di Pisa}\\
\end{center}
\vskip .8truecm
\begin{center}
September, 1995
\end{center}
\vskip 2cm
\begin{center}
To be published in the proceedings of Lattice 95.
\end{center}
\end{titlepage}

\begin{abstract}
By restricting the functional integration to the Regge geometries,
we give the discretized version of the well known path integral
formulation of 2--dimensional quantum gravity in the conformal
gauge. We analyze the role played by diffeomorphisms in the Regge
framework and we give an exact expression for the Faddeev--Popov
determinant related to a Regge surface; such an expression in the
smooth limit goes over to the correct continuum result.
\end{abstract}

While Regge discretized formulation of gravity is well understood at
the classical level, the same cannot be said at the quantum level.
Here the discussion centers mainly about the integration
measure to be adopted in the functional integral and the role
played by diffeomorphisms.

In lattice QCD there exists at the discretized level a well defined
invariance group, i.e.\ the local gauge group which dictates the
unique integration measure; the volume of the gauge group is finite
and in principle no gauge fixing is necessary as we can integrate over
the whole group. With regards to gravity if one wants to maintain the
classical definition of diffeomorphism one must describe the
space--time as a continuous manifold. The Regge approach can be understood
as a description of continuous geometries in which the geometries are
restricted to the ones which are piecewise flat \cite{lee,sork}. From
this point of view one can consider diffeomorphisms in the same way
as in the usual continuous formulation.  In the following we shall
adhere to a Regge theory understood as exactly invariant under the
full group of diffeomorphisms.

With regard to the functional integration in the usual gauge theories
it is performed on the gauge fields and not on the invariants of the
theory; also in gravity we shall stick to the integration over the
metrics and not over some invariants of the metric. In order to do
that a distance among different metrics has to be given and we shall
adopt the usual De-Witt super-metric, which is the unique ultra--local
distance invariant under diffeomorphisms
\begin{equation}
(\delta g^{(1)}, \delta g^{(2)}) = \int \! \sqrt{g} \; \delta
   g^{(1)}_{\mu\nu} \left( g^{\mu\alpha}g^{\nu\beta} +
   g^{\mu\beta}g^{\nu\alpha} + C g^{\mu\nu}g^{\alpha\beta} \right)
   \delta g^{(2)}_{\alpha\beta}.
\end{equation}
Due to the infinite volume of the diffeomorphism group, it is
necessary to introduce a gauge--fixing and the related Faddeev--Popov
determinant has to be taken into account.  Such an approach is the one
which has proven successful in the continuous formulation \cite{poly,alva}
and has been suggested by Jevicki and Ninomiya \cite{jev} in the Regge
approach. In the usual parameterization, the Regge surface in two
dimensions (as in all dimensions) is described by a number of bone
lengths $l_{i}$ which have the meaning of geodesic distances among
points of singular curvature. An infinite family of
$g_{\mu\nu}(x,l_{i})$ will describe such a Regge surface and they will
be related among each other by a diffeomorphism.  We remark that to
choose an element of such equivalence class is not sufficient to
provide a $g_{\mu\nu}$ for each triangular face (in $D=2$) which
correctly describes the metric properties of such a face, e.g.\ by
stating that the metric in the triangle $T_1$ in the figure is given
by
\begin{equation}
g_{1} =
\left(
\begin{array}{cc}
l_{1}^{2} &  \frac{1}{2} (l_{1}^{2} + l_{2}^{2} - l_{3}^{2}) \\
\frac{1}{2} (l_{1}^{2} + l_{2}^{2} - l_{3}^{2}) & l_{2}^{2}
\end{array}
\right).  \label{triang}
\end{equation}
In order to keep the usual meaning of manifold and hence of
diffeomorphism, we must cover our space by open charts (e.g. $C_{1}$
and $C_{2}$) which on the overlap regions must be connected by
transition functions which do not depend on the $l_{i}$ i.e.\ on the
metric \cite{kn}. Under this requirement it is impossible to extend
continuously the metric (\ref{triang}) on the chart $C_{1}$ and the
analogous metric for the triangle $T_{2}$ on the chart $C_{2}$.
Requiring compatibility of the metric on the overlapping regions
severely restricts the consistent choice of the $g_{\mu\nu}$ on each
chart, as we will see below in the case of the sphere topology.
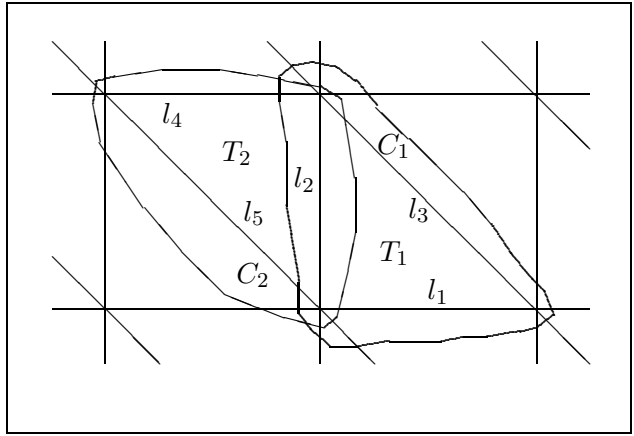
\begin{figure}
\begin{center}
\setlength{\unitlength}{0.012500in}
\begingroup\makeatletter\ifx\SetFigFont\undefined
\gdef\SetFigFont#1#2#3{%
  \ifnum #1<17\tiny\else \ifnum #1<20\small\else
  \ifnum #1<24\normalsize\else \ifnum #1<29\large\else
  \ifnum #1<34\Large\else \ifnum #1<41\LARGE\else
     \huge\fi\fi\fi\fi\fi\fi
  \csname #3\endcsname}%
\fi\endgroup
\begin{picture}(261,180)(39,580)
\thinlines
\put( 58,722){\line( 1, 0){225}}
\put( 58,632){\line( 1, 0){225}}
\put( 80,744){\line( 0,-1){135}}
\put(170,744){\line( 0,-1){135}}
\put(261,744){\line( 0,-1){135}}
\put( 58,654){\line( 1,-1){ 45}}
\put(238,744){\line( 1,-1){ 45}}
\put( 58,744){\line( 1,-1){135}}
\put(148,744){\line( 1,-1){135}}
\multiput(159,734)(-0.42857,-0.35714){15}{\makebox(0.1111,0.7778)
{\SetFigFont{5}{6}{rm}.}}
\put(153,729){\line( 0,-1){ 10}}
\put(153,719){\line( 1,-6){  2.838}}
\put(156,702){\line( 0,-1){ 27}}
\multiput(156,675)(0.09524,-0.57143){22}{\makebox(0.1111,0.7778)
{\SetFigFont{5}{6}{rm}.}}
\multiput(158,663)(0.10000,-0.60000){11}{\makebox(0.1111,0.7778)
{\SetFigFont{5}{6}{rm}.}}
\multiput(159,657)(0.09297,-0.55783){26}{\makebox(0.1111,0.7778)
{\SetFigFont{5}{6}{rm}.}}
\put(161,643){\line( 0,-1){ 13}}
\multiput(161,630)(0.34400,-0.45867){16}{\makebox(0.1111,0.7778)
{\SetFigFont{5}{6}{rm}.}}
\multiput(166,623)(0.42968,-0.35807){20}{\makebox(0.1111,0.7778)
{\SetFigFont{5}{6}{rm}.}}
\put(174,616){\line( 1, 0){ 12}}
\multiput(186,616)(0.56404,0.09401){24}{\makebox(0.1111,0.7778)
{\SetFigFont{5}{6}{rm}.}}
\put(199,618){\line( 1, 0){ 22}}
\multiput(221,618)(0.56491,0.11298){17}{\makebox(0.1111,0.7778)
{\SetFigFont{5}{6}{rm}.}}
\put(230,620){\line( 1, 0){ 10}}
\multiput(240,620)(0.55784,0.09297){26}{\makebox(0.1111,0.7778)
{\SetFigFont{5}{6}{rm}.}}
\multiput(254,622)(0.54298,0.13575){14}{\makebox(0.1111,0.7778)
{\SetFigFont{5}{6}{rm}.}}
\multiput(261,624)(0.45867,0.34400){16}{\makebox(0.1111,0.7778)
{\SetFigFont{5}{6}{rm}.}}
\multiput(268,629)(-0.21053,0.52632){20}{\makebox(0.1111,0.7778)
{\SetFigFont{5}{6}{rm}.}}
\multiput(264,639)(-0.41667,0.41667){13}{\makebox(0.1111,0.7778)
{\SetFigFont{5}{6}{rm}.}}
\multiput(259,644)(-0.36394,0.43672){26}{\makebox(0.1111,0.7778)
{\SetFigFont{5}{6}{rm}.}}
\multiput(250,655)(-0.32000,0.48000){26}{\makebox(0.1111,0.7778)
{\SetFigFont{5}{6}{rm}.}}
\put(242,667){\line(-5, 6){ 11.803}}
\put(230,681){\line(-1, 1){ 22.500}}
\put(207,703){\line(-1, 1){ 13.500}}
\multiput(194,717)(-0.35861,0.43033){25}{\makebox(0.1111,0.7778)
{\SetFigFont{5}{6}{rm}.}}
\multiput(185,727)(-0.44444,0.33333){19}{\makebox(0.1111,0.7778)
{\SetFigFont{5}{6}{rm}.}}
\multiput(177,733)(-0.55556,0.11111){19}{\makebox(0.1111,0.7778)
{\SetFigFont{5}{6}{rm}.}}
\multiput(167,735)(-0.56756,-0.09459){15}{\makebox(0.1111,0.7778)
{\SetFigFont{5}{6}{rm}.}}
\multiput( 75,718)(0.09291,0.55744){17}{\makebox(0.1111,0.7778)
{\SetFigFont{5}{6}{rm}.}}
\multiput( 76,727)(0.57143,0.14286){15}{\makebox(0.1111,0.7778)
{\SetFigFont{5}{6}{rm}.}}
\put( 84,729){\line( 6, 1){ 19.946}}
\put(104,732){\line( 1, 0){ 24}}
\put(128,732){\line( 6,-1){ 15.892}}
\put(144,730){\line( 5,-1){ 25.962}}
\multiput(170,725)(0.49019,-0.29412){19}{\makebox(0.1111,0.7778)
{\SetFigFont{5}{6}{rm}.}}
\put(179,720){\line( 1,-6){  5.513}}
\put(185,687){\line( 0,-1){ 23}}
\put(185,664){\line(-1,-5){  3.192}}
\put(182,648){\line(-1,-4){  5}}
\multiput(177,628)(-0.45455,-0.36364){12}{\makebox(0.1111,0.7778)
{\SetFigFont{5}{6}{rm}.}}
\put(172,624){\line(-4, 1){ 16.941}}
\put(155,628){\line(-5, 2){ 25}}
\put(130,638){\line(-1, 1){ 18}}
\put(112,656){\line(-5, 6){ 15.410}}
\put( 96,674){\line(-2, 3){ 18.462}}
\put( 78,702){\line(-1, 5){  3.192}}
\put( 39,580){\framebox(261,180){}}
\put(215,636){\makebox(0,0)[lb]{\smash{\SetFigFont{18}{6.0}{rm}$l_1$}}}
\put(159,683){\makebox(0,0)[lb]{\smash{\SetFigFont{18}{6.0}{rm}$l_2$}}}
\put(207,670){\makebox(0,0)[lb]{\smash{\SetFigFont{18}{6.0}{rm}$l_3$}}}
\put(104,710){\makebox(0,0)[lb]{\smash{\SetFigFont{18}{6.0}{rm}$l_4$}}}
\put(138,669){\makebox(0,0)[lb]{\smash{\SetFigFont{18}{6.0}{rm}$l_5$}}}
\put(135,643){\makebox(0,0)[lb]{\smash{\SetFigFont{18}{6.0}{rm}$C_2$}}}
\put(194,697){\makebox(0,0)[lb]{\smash{\SetFigFont{18}{6.0}{rm}$C_1$}}}
\put(195,652){\makebox(0,0)[lb]{\smash{\SetFigFont{18}{6.0}{rm}$T_1$}}}
\put(129,694){\makebox(0,0)[lb]{\smash{\SetFigFont{18}{6.0}{rm}$T_2$}}}
\end{picture}
\end{center}
\caption{Triangulation and covering by charts.}
\end{figure}

As done in the 2--dimensional continuum formulation we shall adopt the
conformal gauge fixing $g_{\mu\nu} = e^{2\sigma}
\hat{g}_{\mu\nu}(\tau_{i})$. In the classical papers
\cite{poly,alva}, it has been proved that in such a gauge the
functional integral in 2--dimensional gravity takes the form
\begin{equation}
{\cal{Z}} = \int {\cal{D}} [\sigma] \, d\tau_{i} \;
\sqrt{\frac{{\det}'(L^{\dag}L)}{\det(\Psi_{i}, \Psi_{j})
    \det(\Phi_{a},\Phi_{b})}}
\label{part}
\end{equation}
where
\begin{eqnarray}
(L\xi)_{\mu\nu} = \nabla_{\mu}\xi_{\nu} + \nabla_{\mu}\xi_{\nu}
- g_{\mu\nu} \nabla^{\rho}\xi_{\rho} \nonumber \\
(L^{\dag}h)_{\nu} = -4 \nabla^{\mu} h_{\mu\nu}
\end{eqnarray}
being $\xi_{\mu}$ a vector field and $h_{\mu\nu}$ a symmetric
traceless tensor field. $\tau_{i}$ are the Teichm\"uller parameters
and $\cal{D}[\sigma]$ is the functional integration measure induced by
the distance
\begin{equation}
(\delta \sigma^{(1)}, \delta \sigma^{(2)}) =  \int \sqrt{\hat{g}}
e^{2\sigma} \delta \sigma^{(1)} \delta \sigma^{(2)}
\end{equation}
and $\Psi_{i}$ and $\Phi_{a}$ are respectively the zero modes of
$L$ and $L^{\dag}$. The dependence on
$\sigma$ of the integrand in (\ref{part}) can be factorized in the
expression $e^{-26 S_{L}}$ where
\begin{eqnarray}
S_{L}[\sigma,\hat{g}(\tau_{i})] = \frac{1}{24\pi} \int d^2 x \,
\sqrt{\hat{g}} \; [ \hat{g}^{\mu\nu} \partial_{\mu} \sigma
\partial_{\nu} \sigma + R_{\hat{g}} \sigma ] \label{liouv}
\end{eqnarray}
is the Liouville action.
With the topology of a sphere (to which we shall restrict hereafter)
there are no Teichm\"uller parameters and consequently no zero modes
of $L^{\dag}$.  Instead there are 6 independent solutions of $(L
\Psi)_{\mu\nu} = 0$ which are called conformal Killing vectors, i.e.\
generators of diffeomorphisms which leave unaltered the conformal
structure of the metric. Thus for a universe with spherical topology
eq.(\ref{part}) reduces to
\begin{equation}
{\cal Z}_{\mbox{sp.}}= \int \! {\cal{D}} [\sigma]  \;
\sqrt{\frac{{\det}'(L^{\dag}L)}{\det(\Psi_{i}, \Psi_{j})}} = \int \!
{\cal{D}} [\sigma] \; e^{-26 S_{L}} \label{parts}
\end{equation}
If we try to evaluate $S_{L}$ for a conformal factor describing a
piecewise flat geometry we obtain a divergent result. Nevertheless
$\displaystyle \frac{{\det}' (L^{\dag}L)}{\det (\Psi_{i}, \Psi_{j})}$
can be defined also for a Regge surface by means of the $Z$--function
regularization \cite{chee}. Our main goal will be the computation of
this functional determinant for a Regge surface.

Following a procedure developed by Aurell and Salomonson \cite{aur}
for the scalar Laplace--Beltrami operator on a
piecewise flat two dimensional surface, we derived the exact
expression for such a determinant.  The first point is to
produce a description of a piecewise flat surface with the topology of a
sphere in terms of a conformal factor.  The manifold is described by a
single chart given by the projective plane and the conformal factor
describing a given geometry is unique apart for the action of the
diffeomorphisms generated by the 6 conformal Killing vectors, i.e.\
the $SL(2,C)$ transformations of the complex plane
\begin{equation}
  \omega' = \frac{a\omega + b}{c\omega + d} \; , \; \;  \omega =
  \frac{\omega'd - b}{-c\omega' + a} \; , \;\;  ad - bc = 1
  \label{sl1}
\end{equation}
under which the conformal factor $\sigma$ changes as
\begin{equation}
  \sigma(\omega) \rightarrow \sigma'(\omega') =
  \sigma(\omega(\omega')) - 2 \ln | a - c \omega' |. \label{sl2}
\end{equation}

For a Regge surface whose singularities have location $\omega_{i}$ in
the projective plane and angular aperture $2\pi\alpha_{i}$
($\alpha_{i}=1$ is the plane), the conformal factor takes the
form\footnote{We remark that the apparent singularity along an edge is
only an extrinsic geometry property related to the usual
representation of a Regge surface, while gravity depends
only on the intrinsic geometry.} \cite{foer} $e^{2\sigma} =
e^{2\lambda_{0}} \prod_{i} | \omega -\omega_{i}|^{2(\alpha_{i} -1)}$
which in the neighborhood of $\omega_{i}$ becomes $e^{2\lambda_{i}}
|\omega - \omega_{i}|^{2(\alpha_{i} -1)}$ with $e^{2\lambda_{i}} =
e^{2\lambda_{0}} \prod_{j \neq i} | \omega_{i} - \omega_{j}
|^{2(\alpha_{j} -1)}$.  In the conformal gauge $L$ and $L^{\dag}$
assume a very simple form
\begin{equation}
L = e^{2\sigma} \frac{\partial}{\partial\bar{\omega}} e^{-2\sigma},
\ \ \ L^{\dag} = - e^{-2\sigma} \frac{\partial}{\partial\omega}.
\end{equation}

By means of the $Z$--function regularization we obtain
\begin{equation}
-\ln ({\det}' (L^{\dag}L)) = Z'(s) |_{s=0} =
\gamma_{E} Z(0) +{\mbox{Finite}}_{\epsilon \rightarrow 0}
\int_{\epsilon}^{\infty} \frac{dt}{t} \mbox{Tr}'(e^{-tL^{\dag}L})
\label{zeta}
\end{equation}
where ${\det}'$ and Tr$'$ mean that the zero modes are excluded. The
standard procedure is to compute the variation of $Z'(0)$ under a
variation of the conformal factor: to this end is necessary to know
the constant term in the asymptotic expansion of the heat kernel
$K(x,y,t)$ and $H(x,y,t)$ of the operators $L^{\dag}L$ and
$LL^{\dag}$.  On the other hand the self adjoint extensions of
$L^{\dag}L$ and $LL^{\dag}$ depend on the boundary condition one
imposes on the eigenfunction at the singularities. The choice of
Aurell and Salomonson for the Laplace--Beltrami operator is Dirichlet
boundary conditions. On the other hand the adoption of Dirichlet
boundary condition for the operator $L^{\dag}L$ and $LL^{\dag}$ gives
rise to a result which is not analytic in the angular opening of the
conic singularity and does not agree with the continuum limit. This is
due to the fact that Dirichlet boundary condition are equivalent to
cutting off the tip of the cone. For this reason we considered a
regularization which consists in replacing the tip of the cone by a
segment of sphere or Poincar\'e pseudo--sphere and then letting the
radius of curvature going to zero keeping constant the integrated
curvature \cite{pmppp}.

Such a limiting procedure can be carried through rigorously with the
result that in such a limit new boundary conditions emerge, after
which a well defined integral representation of the heat kernels $K$
and $H$ can be given for $\frac{1}{2} < \alpha < 2$. In particular by
considering the constant terms $c_{0}^{K}=\sum_{i} c_{0\ i}^{K}$ and
$c_{0}^{H}=\sum_{i} c_{0\ i}^{H}$ in the asymptotic expansions of the
trace of $H$ and $K$, we obtain
\begin{eqnarray}
  c_{0\ i}^{K} = \frac{(\alpha_{i} -1)(\alpha_{i} -2)}{2\alpha_{i}} +
  \frac{1-\alpha_{i}^{2}}{12\alpha_{i}} \\
  c_{0\ i}^{H} = \frac{(2\alpha_{i} -1)(2\alpha_{i} -2)}{2\alpha_{i}} +
  \frac{1-\alpha_{i}^{2}}{12\alpha_{i}}
\end{eqnarray}
where $c_{0\ i}^{K}$ and $c_{0\ i}^{H}$ are the contributions of
a single singularity of opening angle $\alpha_{i}$.

We notice that the $c_{0\ i}$ are analytic in $\alpha_{i}$ and
$2(c_{0\ i}^{K} - c_{0\ i}^{H}) = 3 (1-\alpha_{i})$.  Thus for a
generic compact surface without boundary such a relation gives
\begin{equation}
2 (c_{0}^{K} - c_{0}^{H}) = 3 \sum_{i} (1-\alpha_{i} ) = 3 \chi
\end{equation}
being $\chi$ the Euler constant of the surface, in agreement with the
Riemann--Roch theorem applied to $L^{\dag}L$ and $LL^{\dag}$
\cite{alva}.  Performing a variation of the conformal factor we obtain
 from (\ref{zeta})
\begin{eqnarray}
\lefteqn{ - \delta \ln \frac{{\det}'(L^{\dag}L)}{\det(\Psi_{i},
    \Psi_{j})} = \gamma_{E} \delta c_{0}^{K} + \sum_{i} \left\{
    (\delta \lambda_{i} -\lambda_{i} \frac{\delta
    \alpha_{i}}{\alpha_{i}}) [ 4 c_{0\ i}^{K} - 2 c_{0\ i}^{H}] +
    \right. } & \\
& + \left. {\mbox{Finite}}_{\epsilon \rightarrow 0} \left[ 4
    {\displaystyle\frac{\delta \alpha_{i}}{\alpha_{i}}} \int\!dx\, \ln
    (\alpha_{i}|x|) \, K_{\alpha_{i}}(x,x,\epsilon) -2
    {\displaystyle\frac{\delta \alpha_{i}}{\alpha_{i}}} \int\!dx\, \ln
    (\alpha_{i}|x|) \, H_{\alpha_{i}}(x,x,\epsilon) \right] \right\}.
    \nonumber
\end{eqnarray}
A differential of this structure \cite{aur} can be integrated to give
\begin{eqnarray}
\lefteqn{\ln \sqrt{\frac{{\det}'(L^{\dag}L)}{\det(\Psi_{i},
\Psi_{j})}} = } \label{fin} \\
& {\displaystyle
=\frac{26}{12} \left\{ \sum_{i,j\neq i}
\frac{(1-\alpha_{i})(1-\alpha_{j})}{\alpha_{i}} \ln |w_{i} - w_{j}| +
\lambda_{0} \sum_{i} (\alpha_{i} - \frac{1}{\alpha_{i}}) -
\sum_{i} F(\alpha_{i}) \right\}  }   \nonumber
\end{eqnarray}
where $F(\alpha)$ is given by a well defined and convergent
integral representation.  Such a formula has the following appealing
features
\begin{enumerate}

\item It is an exact result giving the value of the $F.P.$ determinant on
  a two dimensional Regge surface.

\item It is invariant under the group $SL(2,C)$ which acts
  on $\omega_{i}$ and $\lambda_{0}$ as
  \begin{eqnarray}
  \lefteqn{\omega_{i} \rightarrow \omega_{i}' =
  \frac{a\omega_{i} +b}{c\omega_{i} + d}} \\
  & \lambda_{0} \rightarrow \lambda_{0}' = \lambda_{0} + \sum_{i}
  (\alpha_{i} -1) \ln |\omega_{i} c + d| \nonumber
  \end{eqnarray}
  and leaves the $\alpha_{i}$ unchanged.

\item In the continuum limit, i.e.\ small angular deficits
  $1-\alpha_{i}$ and dense set of $\omega_{i}$, the first two terms of
  (\ref{fin}) go over to the well know continuum formula
  \begin{equation}
  {\displaystyle
  \frac{26}{96\pi} \left\{\int  dx\,dy \;(\sqrt{g} R)_{x}
  \frac{1}{\Box}(x,y)(\sqrt{g} R)_{y}
  -2 (\ln \frac{A}{\bar{A}})
  \int \: dx\,\sqrt{g} R \right\} }  \label{liouv2}
  \end{equation}
  as can be easily checked, where $A$ is the area $\int dx\,\sqrt{g}$
  and $\bar{A}$ is the area evaluated for $\lambda_{0}=0$. The
  remainder $\sum_{i}
  F(\alpha_{i})$ goes over to a constant topological term.

\item While $\alpha_{i}>0$ with $\sum_{i}(1-\alpha_{i})=2$, the
  $\omega_{i}$ vary without restriction in the complex plane. This as
  pointed out by Foerster \cite{foer} is an advantage over the
  equivalent parameterization of the Regge surface in term of the bone
  length $l_{i}$ where one has to keep into account of a large number
  of triangular inequalities.

\end{enumerate}

We come now to the last piece appearing in (\ref{part}) i.e.\
$\cal{D}[\sigma]$.  Formally one can write
\begin{equation}
{\cal D}[\sigma] =
\sqrt{(\det J)} \, d^{2}\omega_{1} \ldots
d^{2}\omega_{N} \, d\alpha_{1} \ldots d\alpha_{N-1} \, d\lambda_{0}
\label{vol}
\end{equation}
being $J$ the determinant of the matrix $J_{ij}= (
\frac{\partial \sigma}{\partial x_{i}}, \frac{\partial
\sigma}{\partial x_{j}} )$ where $x_{i}$ represent the $3N$
variables $\omega_{1x}, \ldots, \omega_{Nx}, \omega_{1y}, \ldots,
\omega_{Ny}, \alpha_{1}, \ldots, \alpha_{N-1}, \lambda_{0}$. It
is immediately seen that the volume element (\ref{vol}) is invariant under
$SL(2,C)$ as it must, due to the invariance of the conformal gauge
fixing under the 6 conformal Killing vectors. The matrix $J_{ij}$ can
be given a very simple geometrical meaning as follows. Doubling the
number of vertices $x_{i} \rightarrow x'_{i}, x''_{i}$ and defining
$A(x',x'')$ as the area of the two dimensional surface described by
$x'_{i}, x''_{i}$, $J_{ij}$ is given by $J_{ij} = \frac{\partial^{2}
A}{\partial x'_{i} \partial x''_{j}} |_{x'=x''=x}$. Diagonal elements
with indexes $\omega_{ix}$ or $\omega_{iy}$ diverge for positive
curvature (i.e.\ for $1-\alpha_{i} > 0$) and thus the analytic
continuation of $J_{ij}$ has to be considered (as usually done in
conformal theories), which does not break the $SL(2,C)$
invariance. $A(x',x'')$ and the elements $J_{ij}$ are given by
integrals which appeared in the {\em old time}\/ conformal theory
\cite{syma}
and
for which up to now no explicit form is known.  In addition to the
ultra--locality, an interesting feature of $\det (J)$ is that it  vanishes any
time one $\alpha_{i}$ equals 1 (no angular deficit); this is
expected from the fact that for $\alpha_{i}=1$ the value of the position
$\omega_{i}$ is irrelevant in determining the geometry.

{\bf Acknowledgments}: We are grateful to A.\ D'Adda for pointing out to us
reference \cite{foer}.


\begin{thebibliography}{9}


\bibitem{lee}  R.\ Friedberg, T.D.\ Lee, {\it  Nucl.\ Phys.} {\bf
                B242} (1984) 145;
                G.\ Feinberg, R.\ Friedberg, T.D.\ Lee, {\it  Nucl.\
                Phys.} {\bf B245} (1984) 343.

\bibitem{sork}  R.\ Sorkin, {\it Phys.\ Rev.} {\bf D12} (1975) 385;
                {\it Phys.\ Rev.} {\bf D23} (1981) 565.

\bibitem{poly}  A.M.\ Polyakov, {\it Phys.\ Lett.} {\bf 103B} (1984) 207;
                G.\ Moore, P.\ Nelson, {\it Nucl.\ Phys.} {\bf B266}
                (1986) 58; J.\ Polchinski, {\it Comm.\ Math.\ Phys.} {\bf 104}
                (1986) 37.

\bibitem{alva}  O.\ Alvarez, {\it Nucl.\ Phys.} {\bf B216} (1983) 125.

\bibitem{jev}   A.\ Jevicki, M.\ Ninomiya,  {\it Phys.\ Rev.} {\bf
                D33} (1986) 1634.

\bibitem{kn}    S.\ Kobayashi, K.\ Nomizu, {\it Foundations of differential
                geometry}, Interscience Publishers (1963).

\bibitem{chee}  J.\ Cheeger, {\it J.\ Diff.\ Geom.} {\bf 18} (1983) 575.

\bibitem{aur}   E.\ Aurell, P.\ Salomonson, {\it Comm.\ Math.\ Phys.}
                {\bf 165} (1984) 233;
                {\it Further results on
                Functional Determinants of Laplacians in Simplicial
                Complexes}, (hep-th/9405140).

\bibitem{foer}  D.\ Foerster, {\it Nucl.\ Phys.} {\bf B283} (1987) 669;
                {\it Nucl.\ Phys.} {\bf B291} (1987) 813.

\bibitem{pmppp} P.\ Menotti, P.P.\ Peirano {\it Phys.\ Lett.} {\bf
                353B} (1995)  444.

\bibitem{syma}  K.\ Symanzik, {\it Lett.\ Nuovo Cimento} {\bf 3}
                (1972) 734.

\end{thebibliography}
\end{document}